\documentclass[12pt]{article}
\usepackage{amssymb}
\usepackage{amsmath}
\usepackage{IEEEtrantools}
\usepackage{pdflscape}
\usepackage{afterpage}
\usepackage{capt-of}
\usepackage{graphicx}
\usepackage{mathtools}
\usepackage[normalem]{ulem}

\numberwithin{equation}{section}

\newcommand{\be}{\begin{eqnarray}}
\newcommand{\ee}{\end{eqnarray}}
\newcommand{\non}{\nonumber}
\newcommand{\id}{\mathbb{I}}

\newcommand{\tr}{\mathop{\rm tr}\nolimits}
\newcommand{\diag}{\mathop{\rm diag}\nolimits}

\newcommand{\csch}{\mathop{\rm csch}\nolimits}

\newcommand{\mA}{\mathcal{A}}
\newcommand{\mB}{\mathcal{B}}
\newcommand{\mC}{\mathcal{C}}
\newcommand{\mD}{\mathcal{D}}
\newcommand{\refs}{|0\rangle}

\usepackage[usenames,dvipsnames]{color}

\begin{document}

\begin{titlepage}
\strut\hfill UMTG--299
\vspace{.5in}
\begin{center}

{\LARGE Towards the solution\\ 
of an integrable $D_{2}^{(2)}$ spin chain}\\
\vspace{1in}
\large 
Rafael I. Nepomechie \footnote{Physics Department,
            P.O. Box 248046, University of Miami, Coral Gables, FL 
			33124 USA, nepomechie@miami.edu},
Rodrigo A. Pimenta \footnote{Institut Denis-Poisson CNRS/UMR 7013 - Universit\'e de Tours - Universit\'e d'Orl\'eans, 
Parc de Grammont, 37200 Tours, France}${}^{,}$\footnote{CAPES Foundation, Ministry of Education of
Brazil, Brasilia - DF, Zip code 70.040-020, Brazil}${}^{,} $\footnote{Instituto de F\'{i}sica de S\~{a}o Carlos, Universidade de S\~{a}o Paulo, Caixa
Postal 369, 13566-590, S\~{a}o Carlos, SP, Brazil, pimenta@ifsc.usp.br}			
and Ana L. Retore  \footnote{Instituto de F\'{i}sica Te\'{o}rica-UNESP, Rua 
			Dr. Bento Teobaldo Ferraz 271, Bloco II 01140-070, 
			S\~{a}o Paulo, Brazil, ana.retore@unesp.br}
\\[0.8in]
\end{center}

\vspace{.5in}

\begin{abstract}
	Two branches of integrable open quantum-group invariant
	$D_{n+1}^{(2)}$ quantum spin chains are known.  For one branch
	($\varepsilon=0$), a complete Bethe ansatz solution has been
	proposed.  However, the other branch ($\varepsilon=1$) has so far
	resisted solution.  In an effort to address this problem, we
	consider here the simplest case $n=1$.  We propose a Bethe ansatz
	solution, which however is not complete, as it describes only the
	transfer-matrix eigenvalues with odd degeneracy. We also 
	consider a proposal for the missing eigenvalues.
    \end{abstract}

\end{titlepage}

\setcounter{footnote}{0}

\section{Introduction}\label{sec:intro}

Among the non-exceptional trigonometric solutions of the Yang-Baxter 
equation (R-matrices) \cite{Bazhanov:1984gu,
Bazhanov:1986mu, Jimbo:1985ua, Kuniba:1991yd}, the R-matrices associated with $D_{n+1}^{(2)}$
are -- by far -- the most complicated. It is therefore not surprising 
that relatively few results are known about the corresponding integrable 
quantum spin chains. Bethe ansatz solutions for the closed  
$D_{n+1}^{(2)}$ chains with periodic boundary 
conditions were proposed by Reshetikhin \cite{Reshetikhin:1987}. 
Following the pioneering work of Sklyanin \cite{Sklyanin:1988yz}, the 
study of open integrable $D_{n+1}^{(2)}$ chains was initiated in 
\cite{Martins:2000xie}, and was pursued further in 
\cite{Malara:2004bi, Nepomechie:2017hgw}.

New families of solutions of the $D_{n+1}^{(2)}$ boundary Yang-Baxter
equation (K-matrices) were recently proposed in 
\cite{Nepomechie:2018wzp}.  These K-matrices depend on the discrete
parameters $p$ (which can take $n+1$ possible values $p=0, \ldots, n$)
and $\varepsilon$ (which can take two possible values $\varepsilon =
0, 1$), see (\ref{functions}).  
The open spin chains constructed with these K-matrices have
quantum group symmetry corresponding to removing the $p^{th}$ node
from the $D_{n+1}^{(2)}$ Dynkin diagram, namely, $U_{q}(B_{n-p})
\otimes U_{q}(B_{p})$ (for both $\varepsilon = 0, 1$).  These spin
chains also have a $p \leftrightarrow n-p$ duality symmetry.  Bethe
ansatz solutions for the open $D_{n+1}^{(2)}$ spin chains with 
$\varepsilon=0$ (and all the possible values of $p$) were
proposed in \cite{Nepomechie:2017hgw, Nepomechie:2018nvl}.  However,
the open $D_{n+1}^{(2)}$ spin chains with $\varepsilon=1$ have so far resisted
solution.

In an effort to address this problem, we consider here the simplest
case $n=1$; that is, we consider the open $U_{q}(B_{1})$-invariant
$D_{2}^{(2)}$ spin chain with $\varepsilon=1$ and the two possible
values of $p$ (namely, 0 and 1).  This model has potential
applications to black hole physics \cite{RobertsonSaleur}.  We propose a
Bethe ansatz solution that is similar to the one for $\varepsilon=0$;
however, unlike the latter solution, it is {\em not} complete: this
solution describes only the transfer-matrix eigenvalues with {\em odd}
degeneracy. It remains a challenge to account for the eigenvalues with 
even degeneracy, which may be related to a higher symmetry of the 
transfer matrix.

The outline of this paper is as follows. In Sec. \ref{sec:basics}, we 
briefly review the construction of the transfer matrix and list some 
of its useful properties. In Sec. \ref{sec:EV}, we try to
determine the eigenvalues of the transfer matrix with $p=0$. We arrive at a 
compact expression for the eigenvalues (\ref{factored}), (\ref{chi}) 
and corresponding Bethe equations (\ref{BE}), which 
unfortunately do not give all the eigenvalues. However, 
for the eigenvalues which {\em can} be described in this way, we find even simpler 
expressions for the eigenvalues (\ref{square}) and Bethe equations 
(\ref{BEsimple}), which closely resemble those of the XXZ chain. We 
then argue that this Bethe ansatz describes the eigenvalues with odd 
degeneracy. In Sec. \ref{sec:p1}, we consider the case $p=1$. We
consider a proposal for the missing eigenvalues in Sec. 
\ref{sec:BAII}, which is motivated by a preliminary algebraic Bethe 
analysis presented in an appendix. 
Our brief conclusions are in Sec. \ref{sec:conclusion}.

\section{Basics}\label{sec:basics}

In this section, we briefly review the construction of the transfer
matrix for the integrable open $U_{q}(B_{1})$-invariant $D_{2}^{(2)}$
spin chain, with a 4-dimensional vector space at each site.  We also
list some useful properties of this transfer matrix.  We begin by
recalling its two basic building blocks: an R-matrix and a K-matrix.

\subsection{R-matrix}

For the $16 \times 16$ R-matrix $R(u)$, we 
use the expression for the $D_{n+1}^{(2)}$ R-matrix in the 
fundamental (vector) representation given in Appendix A of 
\cite{Nepomechie:2017hgw} with $n=1$. This R-matrix depends on the anisotropy 
parameter $\eta$. In addition to the Yang-Baxter equation, 
it satisfies the unitarity relation
\be
R_{12}(u)\, R_{21}(-u) = \zeta(u) \,, \qquad \zeta(u) = 16 
\sinh^{2}(u+2\eta)\, \sinh^{2}(u-2\eta) \,,
\label{unitarity}
\ee
and the crossing-unitarity relation
\be
M^{-1}_{1}\, R_{12}(-u-2\rho)^{t_{1}}\, M_{1}\, R_{21}(u)^{t_{1}} = 
\zeta(u+\rho) \,, \qquad \rho = -2\eta \,,
\ee
where $M$ is the diagonal $4 \times 4$ matrix
\be
M = \diag\left(e^{2\eta}\,, 1\,, 1\,, e^{-2\eta} \right) \,.
\ee 

\subsection{K-matrix}

For the right K-matrix $K^{R}(u)$, we take \cite{Nepomechie:2018wzp}
{\small
\be 
K^{R}(u) =  
\left( \begin{array}{c|c|cc|c|c}
k_{-}(u)\, \id_{p \times p} & & & & & \\
\hline
& g(u)\, \id_{(n-p) \times (n-p)} & & & &\\
\hline
& & k_{1}(u) & k_{2}(u) & &\\
& & k_{2}(u) & k_{1}(u) & & \\
\hline
& & & & g(u)\, \id_{(n-p) \times (n-p)} &  \\
\hline
& & & & & k_{+}(u)\, \id_{p \times p}
\end{array} \right) \,,
\label{KR}
\ee}
with
\begin{align}
k_{\mp}(u) &= e^{\mp 2u} \,, \non \\
g(u) &= \frac{\cosh(u-(n-2p)\eta + 
\frac{i\pi}{2}\varepsilon)}{\cosh(u+(n-2p)\eta - \frac{i\pi}{2}\varepsilon)} \,, \non \\
k_{1}(u)  &= \frac{\cosh(u) \cosh((n-2p)\eta + \frac{i\pi}{2}\varepsilon)}
{\cosh(u+(n-2p)\eta)+\frac{i\pi}{2}\varepsilon)} 
\,, \non \\
k_{2}(u)  &= -\frac{\sinh(u) \sinh((n-2p)\eta + \frac{i\pi}{2}\varepsilon)}
{\cosh(u+(n-2p)\eta + \frac{i\pi}{2}\varepsilon)} \,,
\label{functions}
\end{align}
see also \cite{Martins:2000xie}.
Since we restrict our attention here to $D_{2}^{(2)}$, which corresponds to 
$n=1$, the matrix $K^{R}(u)$ is $4 \times 4$.
There are two possible values of $p$ (namely, 0 and 1), and we now 
set $p=0$. (We consider the $p=1$ case in Sec. \ref{sec:p1}.) As emphasized in the 
Introduction, we focus in this paper on the case $\varepsilon = 1$. 

For the left K-matrix $K^{L}(u)$, we take \cite{Nepomechie:2018wzp}
\be
K^{L}(u) = K^{R}(-u-\rho)^{t}\, M \,,
\label{KLfundam}
\ee 
so that the transfer matrix has quantum-group symmetry.

\subsection{Transfer matrix}

The transfer matrix $t(u)$ for an open integrable quantum spin chain of length $N$
is given by \cite{Sklyanin:1988yz, Nepomechie:2018wzp}
\be
t(u) = \tr_a K^{L}_{a}(u)\, T_a(u; \{\theta_{j}\})\,  K^{R}_{a}(u)\, 
\widehat{T}_a(u; \{\theta_{j}\}) \,, 
\label{transfer}
\ee
where the monodromy matrices with inhomogeneities $\{ \theta_{1}\,, 
\ldots\,, \theta_{N} \}$ are given by
\begin{align}
T_a(u; \{\theta_{j}\}) &= R_{aN}(u-\theta_{N})\ R_{a 
N-1}(u-\theta_{N-1})\ \cdots R_{a1}(u-\theta_{1}) \,,  \non \\
\hat T_a(u; \{\theta_{j}\})  &= R_{1a}(u+\theta_{1})\ \cdots R_{N-1 a}(u+\theta_{N-1})\ 
R_{Na}(u+\theta_{N}) \,,
\label{monodromyinhomo}
\end{align}
and the trace in (\ref{transfer}) is over the (4-dimensional) auxiliary 
space. 

\subsection{Properties of the transfer 
matrix}\label{subsec:transferprops}

By construction, the transfer matrix satisfies the 
commutativity property
\be
\left[ t(u) \,, t(v) \right] = 0 \,.
\ee
The transfer matrix also obeys the functional relations (see 
\cite{Nepomechie:2017hgw} and references therein)
\be
t(\theta_{j})\, t(\theta_{j} +2\eta) 
= \Delta(\theta_{j}) \,, \qquad 
j = 1, \ldots, N\,,  \label{fr1}
\ee   
where
\begin{align}
h^{(R)}(u) &= 2^{9} e^{-2\eta} \cosh(u-2\eta)\, \cosh^{2}(u-\eta)\, 
\sinh(u-5\eta)\, \sinh(u-4\eta)\,  \non \\
& \qquad \times \sinh^{2}(2u-6\eta)\, \sinh(2u-4\eta)\, \sinh(u-\eta) \,, \\
h^{(L)}(u) &= 2^{7} e^{2\eta} \csch(u-7\eta)\cosh(u-6\eta)\, \sinh(u-4\eta) \non \\
& \qquad \times \sinh^{2}(2u-8\eta)\, \sinh^{2}(2u-4\eta)\, 
\sinh(2u-12\eta)\, \sinh(u-3\eta) \,, \\
h(u) &= h^{(L)}(u)\, h^{(R)}(u) \prod_{k=1}^{N} \zeta(u-4\eta + 
\theta_{k})\, \zeta(u-4\eta - \theta_{k})\,, \\
\Delta(u) &= \frac{h(u+4\eta)}{\zeta(2u)\, \zeta(2u+2\eta)\, 
\zeta(2u+4\eta)} \,,
\end{align}
and $\zeta(u)$ is defined in (\ref{unitarity}).
The transfer matrix also has $i \pi$ periodicity
\be
t(u) = t(u + i \pi)\,, \label{periodicity}
\ee
as well as crossing symmetry
\be
t(u) = t(-u+2\eta) \,.
\label{crossing}
\ee
Finally, the transfer matrix has
the particular value 
\be
	\lim_{u\rightarrow 0} \frac{t(u)}{\sinh u} \Big\vert_{\{ \theta_{j} \} = 0}
	= 2^{4N} 
	\sinh^{4N-3}(2\eta) \sinh^{2}(4\eta) \sinh(\eta) \csch(3\eta) \,.
	\label{particular}
\ee

\section{Eigenvalues of the transfer matrix}\label{sec:EV}

This section, which is devoted to determining the transfer-matrix
eigenvalues, contains most of our new results. We first show in Sec. 
\ref{subsec:eigprops} that the transfer-matrix properties listed in Sec. \ref{subsec:transferprops} 
do not suffice to determine the eigenvalues. We then formulate in Sec. 
\ref{subsec:conjecture} a conjecture for the eigenvalues, which is 
developed further in Sec. \ref{subsec:BAI}.

\subsection{Properties of the eigenvalues}\label{subsec:eigprops}

Let $\Lambda(u)$ denote the eigenvalues of the transfer matrix $t(u)$. It follows from 
the transfer-matrix properties (\ref{fr1}) - (\ref{particular}) that the 
eigenvalues satisfy similar properties:
\be
\Lambda(\theta_{j})\, \Lambda(\theta_{j} +2\eta) 
= \Delta(\theta_{j}) \,, \qquad 
j = 1, \ldots, N\,,  \label{Lfr1}
\ee 
\be
\Lambda(u) = \Lambda(u + i \pi)\,, \label{Lperiodicity}
\ee
\be
\Lambda(u) = \Lambda(-u+2\eta)\,,
\label{Lcrossing}
\ee
\be
	\lim_{u\rightarrow 0} \frac{\Lambda(u)}{\sinh u}\Big\vert_{\{ \theta_{j} \} = 0}
	= 2^{4N} 
	\sinh^{4N-3}(2\eta) \sinh^{2}(4\eta) \sinh(\eta) \csch(3\eta) \,.
	\label{Lparticular}
\ee
Moreover, the eigenvalues have the asymptotic behavior
\be
	\Lambda(u)  \sim  2 e^{\pm 4 N (u- \eta)} \left[\cosh( 
	4\eta(N-m+\tfrac{1}{2})) + 1 \right]\quad \mbox{   for } u 
	\rightarrow \pm \infty \,, 
\label{Lasymptotic}
\ee
where $m$ is a non-negative integer. 

In order to proceed further, it is convenient to consider rescaled 
eigenvalues $\lambda(u)$, defined such that
\be
\Lambda(u)  = \phi(u)\, \lambda(u) \,, 
\qquad \phi(u) = \frac{\sinh(u) \sinh(u-2\eta)}{\sinh(u+\eta) \sinh(u-3\eta)} \,.
\label{rescaled}
\ee
The rescaled eigenvalues $\lambda(u)$ do not have any poles for finite $u$, and 
do not have zeros at $u=0, 2\eta$.
The rescaled eigenvalues have the properties
\be
\lambda(\theta_{j})\, \lambda(\theta_{j} +2\eta) = 
\frac{\Delta(\theta_{j})}{\phi(\theta_{j})\, \phi(\theta_{j}+2\eta)} \,,  \qquad 
j = 1, \ldots, N\,,  \label{fffr1}
\ee  
and
\begin{align}
\lambda(u) & = \lambda(u + i \pi)\,, \label{periodicity2}\\
\lambda(u) & = \lambda(-u+2\eta)\,, \label{crossing2} \\
\lambda(0) \Big\vert_{\{ \theta_{j} \} = 0}
&= 2^{4N} 
	\sinh^{4N-4}(2\eta) \sinh^{2}(4\eta) \sinh^{2}(\eta)\,,
	\label{part2} \\
\lambda(u)  & \sim  2 e^{\pm 4 N (u- \eta)} \left[\cosh( 
	4\eta(N-m+\tfrac{1}{2})) + 1 \right]\quad \mbox{   for } u 
	\rightarrow \pm \infty \,. \label{asym2} 
\end{align}	

The periodicity (\ref{periodicity2}) and asymptotic behavior (\ref{asym2})
imply that the eigenvalues have the form
\be 
\lambda(u) = \sum_{k=-2N}^{2N}\lambda_{k} e^{2k u} \,,
\label{expansionlambda}
\ee
where $\lambda_{k}$ are $u$-independent coefficients, of which there are 
$4N+1$. However, the crossing symmetry (\ref{crossing2}) relates the coefficients $\lambda_{k>0}$ to 
$\lambda_{k<0}$. Hence, there are $2N+1$ independent coefficients for $\lambda(u)$.

The functional relations (\ref{fffr1}) provide $N$ 
constraints. The asymptotic behavior (\ref{asym2}) provides one constraint
(the behavior at $-\infty$ follows from the behavior at $+\infty$ 
together with crossing symmetry),
and (\ref{part2}) provides one
more constraint, for a total of only $N+2$ constraints.
Therefore, for $N>1$, these constraints do {\em not} suffice to determine 
$\lambda(u)$. 

We have tried to obtain additional constraints by formulating
functional relations involving fused transfer matrices, as in e.g.
\cite{Hao:2014fha, Li:2018xrb}.  However, this introduces even more
unknown coefficients (to describe the eigenvalues of the fused
transfer matrices, similarly to (\ref{expansionlambda})), and does not seem to help.

In the next subsection, we conjecture an expression for $\lambda(u)$ 
that is compatible with the above constraints.

\subsection{Formulating a conjecture for 
$\lambda(u)$}\label{subsec:conjecture}

In view of the result for $\varepsilon=0$ \cite{Nepomechie:2017hgw,
Nepomechie:2018nvl}, let us assume that the rescaled eigenvalues 
$\lambda(u)$ have the form
\be
	\lambda(u)  = Z_{1}(u) + Z_{2}(u) + Z_{3}(u) + Z_{4}(u) \,, 
	\label{Ansatz} 
\ee
where
\begin{align}
	Z_{1}(u) & =  a(u)\, \frac{Q(u+\eta)\, Q(u+\eta+i \pi)}{Q(u-\eta)\, 
	Q(u-\eta+i \pi)} 
	\prod_{k=1}^{N}16\sinh^{2}(u-\theta_{k}-2\eta)\sinh^{2}(u+\theta_{k}-2\eta)\,, \non \\
	Z_{2}(u) & = b(u)\, \frac{Q(u-3\eta)\, Q(u+\eta+i \pi)}{Q(u-\eta)\, 
	Q(u-\eta+i \pi)} \non \\
	& \quad \times \prod_{k=1}^{N}16\sinh(u-\theta_{k})\sinh(u-\theta_{k}-2\eta)\sinh(u+\theta_{k})\sinh(u+\theta_{k}-2\eta)
	\,, \non \\
	Z_{3}(u) & = b(-u+2\eta)\, \frac{Q(u+\eta)\, Q(u-3\eta+i \pi)}{Q(u-\eta)\, 
	Q(u-\eta+i \pi)} \non \\
	& \quad \times \prod_{k=1}^{N}16\sinh(u-\theta_{k})\sinh(u-\theta_{k}-2\eta)\sinh(u+\theta_{k})\sinh(u+\theta_{k}-2\eta)
	\,, \non \\
	Z_{4}(u) & = a(-u+2\eta)\,\frac{Q(u-3\eta)\, Q(u-3\eta+i \pi)}{Q(u-\eta)\, 
	Q(u-\eta+i \pi)}  \prod_{k=1}^{N}16\sinh^{2}(u-\theta_{k})\sinh^{2}(u+\theta_{k})\,, 
	\label{Zs}
\end{align}	
and
\be
Q(u) = \prod_{j=1}^{m}\sinh(\tfrac{1}{2}(u- u_{j}))\, 
\sinh(\tfrac{1}{2}(u+ u_{j})) \,,
\label{Q}
\ee
where the functions $a(u)$ and $b(u)$ are still to be determined. 

The function $a(u)$ can readily be seen to be given by
\be
	a(u) = \frac{\cosh^{2}(u-2\eta)}{\cosh^{2}(u-\eta)} \,, 
\label{afunc}	
\ee
either from the functional relation (\ref{fffr1}), or by explicitly computing the 
reference-state eigenvalue for small values of $N$ and with
values of the inhomogeneities chosen such that 
only $Z_{1}(u)$ is nonzero, as explained in 
detail in \cite{Nepomechie:2018nvl}. Note that 
$a(u)$ (\ref{afunc}) has a double-pole at $u=\eta + \frac{i \pi}{2}$.

The function $b(u)$ must have the same double-pole as $a(u)$ in order 
for $\lambda(u)$ (\ref{Ansatz}) to be analytic. We therefore set
\be
b(u) =  \frac{c(u)}{\cosh^{2}(u-\eta)}  \,,
\label{bform}
\ee
where $c(u)$ is finite at $u=\eta + \frac{i \pi}{2}$. The function $b(u)$ 
must also satisfy
\be
b(u) + b(-u+2\eta) = \frac{2\cosh(u)\cosh(u-2\eta)}{\cosh^{2}(u-\eta)} 
\ee
in order to ensure that $\lambda(u)$ is correct for the reference 
state, for which $Q(u)=1$. Therefore, $c(u)$ satisfies
\be
c(u) + c(-u+2\eta) = 2\cosh(u)\cosh(u-2\eta) \,.
\label{c1}
\ee
The condition that the residue of $\lambda(u)$ (\ref{Ansatz}) at the double-pole 
vanishes implies
\be
c'(\eta \pm \frac{i \pi}{2}) = 0 \,,
\label{c2}
\ee
where prime denotes differentiation. Finally, let us assume that $b(u)$ (and 
therefore also $c(u)$) is $i \pi$ periodic \footnote{The weaker 
assumption
$$ b(-u+2\eta) = b(u + i \pi) \,, \qquad b(u) = b(u + 2 i \pi) \,, $$
is also compatible with the $i \pi$ periodicity of $\lambda(u)$ (\ref{periodicity2}), 
and leads to 
$$ b(u) = \frac{\cosh(u)\cosh(u-2\eta)}{\cosh^{2}(u-\eta)} + \beta 
\frac{\sinh(u-\eta)}{\cosh^{2}(u-\eta)} \,, $$
where $\beta$ is a free parameter, cf. (\ref{bfunc}). However, 
even for $N=2$, we cannot find any value of $\beta$ for which (\ref{Ansatz}) 
gives all the eigenvalues.}
\be
b(u) = b(u + i \pi)\,,
\ee
and has the asymptotic behavior
\be
\lim_{u\rightarrow \pm \infty} b(u) = \mbox{  finite }
\ee
(which is compatible with (\ref{asym2})), which imply that $c(u)$ has the 
form
\be
c(u) = \sum_{k=-1}^{1}c_{k} e^{2k u}  \,,
\ee
where $c_{k}$ are independent of $u$.
The constraints (\ref{c1}) and (\ref{c2}) then uniquely determine 
$c(u)$ to be given by
\be
c(u) = \cosh(u)\cosh(u-2\eta) \,.
\ee
It follows from (\ref{bform}) that $b(u)$ is given by
\be
b(u) = b(-u+2\eta) = \frac{\cosh(u)\cosh(u-2\eta)}{\cosh^{2}(u-\eta)}  \,.
\label{bfunc}
\ee

In summary, we conjecture that the rescaled eigenvalues $\lambda(u)$ 
are given by (\ref{Ansatz}) and (\ref{Zs}), where $Q(u)$, $a(u)$ and $b(u)$ given by 
(\ref{Q}), (\ref{afunc}) and (\ref{bfunc}), respectively.
This ansatz satisfies all the constraints (\ref{fffr1}) - (\ref{part2}).  

\subsection{Bethe ansatz}\label{subsec:BAI}

We observe that this expression for $\lambda(u)$ can be factored as 
follows \footnote{For the case $\varepsilon=0$ 
\cite{Nepomechie:2017hgw, Nepomechie:2018nvl}, and 
presumably also for the periodic chain \cite{Reshetikhin:1987},
such a factorization is possible for all the eigenvalues, hence it may hold at the level 
of the transfer matrix.}
\be
\lambda(u) = \chi(u)\, \chi(u + i\pi) \,,
\label{factored}
\ee
where $\chi(u)$ is defined by
\begin{align}
\chi(u) & = \frac{\cosh(u-2\eta)}{\cosh(u-\eta)} \frac{Q(u+\eta)}{Q(u-\eta)}
\prod_{k=1}^{N}4\sinh(u-\theta_{k}-2\eta)\sinh(u+\theta_{k}-2\eta) 
\non \\
& +
\frac{\cosh(u)}{\cosh(u-\eta)} \frac{Q(u-3\eta)}{Q(u-\eta)}
\prod_{k=1}^{N}4\sinh(u-\theta_{k})\sinh(u+\theta_{k}) \,,
\label{chi}
\end{align}
which satisfies $\chi(-u+2\eta) = \chi(u)$.
The requirement that the residues of $\chi(u)$ vanish at $u=u_{j} + \eta$ leads to the Bethe 
equations for $\{ u_{1}, \ldots, u_{m} \}$
\begin{align}
\MoveEqLeft \prod_{l=1}^{N} \frac{\sinh(u_{j} -\theta_{l} + 
\eta)}{\sinh(u_{j} -\theta_{l} - \eta)}
\frac{\sinh(u_{j} + \theta_{l} + 
\eta)}{\sinh(u_{j} + \theta_{l} - \eta)} \non \\
& = \frac{\sinh(u_{j}+\eta)\, \cosh(u_{j}-\eta)}
{\sinh(u_{j}-\eta)\, \cosh(u_{j}+\eta)}
\prod_{k=1;\, k \ne j}^{m} 
\frac{\sinh(\tfrac{1}{2}(u_{j} - u_{k}) + \eta)}
{\sinh(\tfrac{1}{2}(u_{j} - u_{k}) - \eta)}
\frac{\sinh(\tfrac{1}{2}(u_{j} + u_{k}) + \eta)}
{\sinh(\tfrac{1}{2}(u_{j} + u_{k}) - \eta)} \,, \non \\
& \qquad\qquad\qquad\qquad\qquad\qquad\qquad\qquad j = 1, \ldots, m\,.
\label{BE}
\end{align}

Unfortunately, this Bethe ansatz solution is {\em not} complete: we 
have checked numerically for small values of $N$ that this solution gives some, but not all, 
of the transfer-matrix eigenvalues. However, for every 
eigenvalue that we {\em do} find, the number of Bethe roots ($m$) is 
even, and all the Bethe roots come in pairs separated by exactly $i \pi$
\be
\{ u_{j}\,, u_{j}  + i \pi \}\,, \qquad j = 1\,, \ldots\,, 
\frac{m}{2} \,. 
\label{pairs}
\ee 

Assuming that the Bethe roots always form pairs of the form 
(\ref{pairs}), then the Q-function (\ref{Q}) becomes (up to an 
irrelevant overall factor) 
\be
Q(u) = \prod_{j=1}^{\frac{m}{2}}\sinh(u- u_{j})\, \sinh(u+ u_{j}) \,,
\label{QQ}
\ee
and therefore $Q(u)$ becomes $i\pi$ periodic
\be
Q(u) = Q(u + i\pi) \,.
\ee 
It follows that $\chi(u)$ (\ref{chi}) also becomes $i\pi$ periodic, and therefore
$\lambda(u)$ (\ref{factored}) becomes a perfect square
\be
\lambda(u) = \chi(u)^{2} \,.
\label{square}
\ee
The requirement that the residues of $\chi(u)$ vanish at $u=u_{j} + 
\eta$ now leads to the simplified Bethe equations 
\begin{align}
\MoveEqLeft \prod_{l=1}^{N} \frac{\sinh(u_{j} -\theta_{l} + 
\eta)}{\sinh(u_{j} -\theta_{l} - \eta)}
\frac{\sinh(u_{j} + \theta_{l} + 
\eta)}{\sinh(u_{j} + \theta_{l} - \eta)} \non \\
& = \frac{\sinh(u_{j}+\eta)}
{\sinh(u_{j}-\eta)}
\prod_{k=1;\, k \ne j}^{\frac{m}{2}} 
\frac{\sinh(u_{j} - u_{k} + 2\eta)}
{\sinh(u_{j} - u_{k} - 2\eta)}
\frac{\sinh(u_{j} + u_{k} + 2\eta)}
{\sinh(u_{j} + u_{k} - 2\eta)} \,, \non \\
& \qquad\qquad\qquad\qquad\qquad\qquad\qquad\qquad j = 1, \ldots, 
\frac{m}{2}\,.
\label{BEsimple}
\end{align}
Interestingly, these Bethe equations are similar to those for the 
spin-1/2 XXZ chain.

We have solved the simplified Bethe equations (\ref{BEsimple}) with all 
$\theta_{l}=0$
numerically (for
some generic value of anisotropy parameter $\eta$) for $N=1, 2,
\ldots, 5$; we have then computed the corresponding eigenvalues (for some generic
value of spectral parameter $u$) using (\ref{rescaled}), (\ref{chi})
and (\ref{square}); and we have compared with the eigenvalues obtained by
direct diagonalization of the transfer matrix (\ref{transfer}).  The
results are summarized in Tables \ref{table:N1} - \ref{table:N5}.  For
a given value of $N$, each table reports the degeneracy (the number of
times a given eigenvalue appears), the multiplicity (the number of
times a given degeneracy appears), and $m$ (twice the number of Bethe
roots of the simplified Bethe equations (\ref{BEsimple}) that are
needed to describe an eigenvalue with the given degeneracy).  A
question mark (?)  means that an eigenvalue with the given degeneracy
cannot be described by this Bethe ansatz.  For example, from
Table \ref{table:N2}, we can see that for $N=2$, there is one
eigenvalue with degeneracy 5 which corresponds to the reference state
($m=0$); there are two eigenvalues with degeneracy 1 which are each
described by 1 Bethe root ($m=2$); there is one eigenvalue with
degeneracy 3 which is described by 2 Bethe roots ($m=4$); and there is
one eigenvalue with degeneracy 6 which cannot be described by this
Bethe ansatz. Note that $m$ takes even values from 0 to $2N$.
(We do not report the actual Bethe roots and eigenvalues
in order to avoid having prohibitively large tables.)

An inspection of these tables shows that our Bethe ansatz describes 
all the eigenvalues with {\em odd} degeneracy, but does not describe 
any of the eigenvalues with even degeneracy. We conjecture that this 
is true for generic values of $\eta$ and for all values of $N$.

For a given value of $N$, let ${\cal N}_{odd}$ and ${\cal N}_{even}$
denote the total number of eigenvalues (given by the product 
degeneracy $\times$ multiplicity)
with odd and even degeneracy, respectively.  Clearly,
\be
{\cal N}_{odd} + {\cal N}_{even} = 4^{N} \,.
\ee
From Tables \ref{table:N1}-\ref{table:N5}, we can see that the fraction of eigenvalues with odd 
degeneracy rapidly decreases as $N$ increases, as summarized in 
Table \ref{table:oddfraction}.

\begin{table}[h!]
\centering
\begin{tabular}{|c|ccccc|}
\hline
$N$ & 1 & 2 & 3 & 4 & 5 \\   
\hline
 ${\cal N}_{odd}/4^{N}$ & 1  & 0.625 & 0.375 & 0.210938 & 0.117188 \\
\hline
\end{tabular}
\caption{Fraction of eigenvalues with odd 
degeneracy}\label{table:oddfraction}
\end{table}

We expect that the ``missing'' eigenvalues (i.e., the eigenvalues 
with even degeneracy, which cannot be described by this Bethe ansatz) 
{\em cannot} be expressed as perfect squares, as in (\ref{square}). We have 
verified this for $N=2$, in which case all the eigenvalues 
can be explicitly computed as functions of $u$ and $\eta$.

\subsubsection{Degeneracies and symmetries}\label{subsubsec:degensym}

On the 
basis of $U_{q}(B_{1})$ symmetry alone, one would expect that every 
eigenvalue of the transfer matrix has odd degeneracy 
\cite{Nepomechie:2017hgw, Nepomechie:2018wzp}. For example,
for $N=2$:
\be
\left({\bf 3} \oplus {\bf 1}\right)^{\otimes 2} 
= 2\cdot {\bf 1} \oplus  3 \cdot{\bf 3} \oplus  {\bf 5}  \,;
\ee
and for $N=3$:
\be
\left({\bf 3} \oplus {\bf 1}\right)^{\otimes 3} 
 = 5 \cdot {\bf 1}  \oplus  9 \cdot {\bf 3} 
\oplus 5 \cdot {\bf 5} \oplus {\bf 7} \,.
\ee
However, we can easily see from Tables \ref{table:N2} and \ref{table:N3} that 
the actual degeneracies are {\em higher}: for $N=2$, one pair of ${\bf 3}$'s 
becomes degenerate (giving a 6-fold degenerate eigenvalue); 
and for $N=3$, two pairs of ${\bf 5}$'s 
become degenerate (giving two 10-fold degenerate eigenvalues), 
three pairs of ${\bf 3}$'s become degenerate (giving three 6-fold degenerate 
eigenvalues), and one pair of ${\bf 1}$'s 
becomes degenerate (giving a 2-fold degenerate eigenvalue).

We have conjectured in \cite{Nepomechie:2017hgw, Nepomechie:2018wzp} 
that these higher (even) degeneracies are due to an additional 
symmetry of the transfer matrix that {\em doubles} the degeneracy of 
certain eigenvalues, for both $\varepsilon=0$ and $\varepsilon=1$.
Indeed, for $N=2$, we have explicitly 
constructed an involutory matrix that maps one ${\bf 3}$ to another ${\bf 3}$,
commutes with all the $U_{q}(B_{1})$ generators, and commutes with 
$t(u)$. However, an extension of this construction to 
general values of $N$ is still not known.

Our new observation here is that the ``missing'' eigenvalues are
precisely those that would become degenerate as the result of this
additional symmetry.

\section{The case $p=1$}\label{sec:p1}

The results discussed so far in Secs. \ref{subsec:transferprops} and \ref{sec:EV} are 
for $p=0$. We now consider the case $p=1$. To this end, it is 
convenient to now change notation so that the dependence 
on $p$ becomes manifest, e.g. $K^{R, L}(u) \mapsto K^{R, L}(u, p)$,
$t(u) \mapsto t(u, p)$, $\Lambda(u) \mapsto \Lambda(u, p)$,  etc.
In particular, the result (\ref{rescaled}) becomes
\be
\Lambda(u, 0) = \phi(u, 0)\, \lambda(u) \,, 
\qquad \phi(u, 0) = \frac{\sinh(u) \sinh(u-2\eta)}{\sinh(u+\eta) \sinh(u-3\eta)} \,.
\label{rescaledp0}
\ee
For $p=1$, we obtain in a similar way 
\be
\Lambda(u, 1) = \phi(u, 1)\, \lambda(u) \,, 
\qquad \phi(u, 1) = \frac{\sinh(u) \sinh(u-2\eta)}{\sinh^{2}(u-\eta)} \,,
\label{rescaledp1}
\ee
where $\lambda(u)$ is again given by 
(\ref{Ansatz}), (\ref{factored}), etc. The Bethe equations are 
therefore also the same as before. In other words, only the 
overall factor changes.

This result is consistent with the $p\leftrightarrow n-p$ duality 
symmetry that was mentioned in the Introduction. Indeed, the transfer 
matrix has the symmetry \cite{Nepomechie:2018wzp}
\be
{\cal U}\, t(u,p)\, {\cal U}^{-1} = f(u,p)\, t(u,n-p)  \,, 
\label{duality}
\ee 
where ${\cal U}$ is a certain operator acting in the quantum space, 
and $f(u,p)$ is a scalar function given by
\be
f(u,p) = f^{L}(u,p)\, f^{R}(u,p) \,,
\label{ffunction}
\ee 
with
\begin{align}
f^{R}(u,p) &= \frac{\cosh(u-(n-2p)\eta + \frac{i\pi}{2}\varepsilon)}
{\cosh(u+(n-2p)\eta - \frac{i\pi}{2}\varepsilon)} \,,  \non \\
f^{L}(u,p) &= 
\frac{\cosh(u-(n+2p)\eta + \frac{i\pi}{2}\varepsilon)}
{\cosh(u-(3n-2p)\eta - \frac{i\pi}{2}\varepsilon)}
\,.
\end{align}
It follows that the corresponding eigenvalues are related by
\be
\Lambda(u,p)  = f(u,p)\, \Lambda(u,n-p) \,.
\label{duality2}
\ee
Substituting (\ref{rescaledp0}) and (\ref{rescaledp1}) into 
(\ref{duality2}) with $n=p=1$ leads to the constraint
\be
f(u,1) = \frac{\phi(u, 1)}{\phi(u, 0)} = \frac{\sinh(u+\eta) 
\sinh(u-3\eta)}{\sinh^{2}(u-\eta)} \,,
\ee
which is indeed consistent with (\ref{ffunction}) for $\varepsilon=1$.

\section{An ansatz for the missing eigenvalues?}\label{sec:BAII}

Let us now consider the following ansatz for the ``missing'' eigenvalues 
with $p=0$
\be
	\lambda(u)  = Z_{1}(u) - Z_{2}(u) - Z_{3}(u) + Z_{4}(u) + Z_{5}(u)\,, 
	\label{tAnsatz} 
\ee
where the functions $Z_{1}(u), \ldots,  Z_{4}(u)$ are given as before 
by (\ref{Zs}), and
\begin{align}
	Z_{5}(u) = 4b(u)\, 
	\prod_{k=1}^{N}16\sinh(u-\theta_{k})\sinh(u-\theta_{k}-2\eta)\sinh(u+\theta_{k})\sinh(u+\theta_{k}-2\eta)\,.
	\label{tZs}
\end{align}	
This ansatz
is very similar to the previous one (\ref{Ansatz}), except for some 
signs and the shift of
all the eigenvalues by $Z_{5}(u)$, which does not 
depend on the Q-function. This ansatz also satisfies the constraints 
(\ref{fffr1})-(\ref{part2}). It is motivated by a 
preliminary algebraic Bethe ansatz analysis, which is presented in 
Appendix \ref{sec:ABA}.

The expression (\ref{tAnsatz})-(\ref{tZs}) for the eigenvalues, up to the shift, can be factored as follows
\be
\lambda(u) = \tilde{\chi}(u)\, \tilde{\chi}(u + i\pi) + 
Z_{5}(u) \,,
\label{tfactored}
\ee
where $\tilde{\chi}(u)$ is defined by
\begin{align}
\tilde{\chi}(u) & = \frac{\cosh(u-2\eta)}{\cosh(u-\eta)} 
\frac{Q(u+\eta + i \pi)}{Q(u-\eta)}
\prod_{k=1}^{N}4\sinh(u-\theta_{k}-2\eta)\sinh(u+\theta_{k}-2\eta) 
\non \\
& -
\frac{\cosh(u)}{\cosh(u-\eta)} \frac{Q(u-3\eta + i \pi)}{Q(u-\eta)}
\prod_{k=1}^{N}4\sinh(u-\theta_{k})\sinh(u+\theta_{k}) \,,
\label{tchi}
\end{align}
which satisfies $\tilde{\chi}(-u+2\eta) = -\tilde{\chi}(u)$.
Requiring that the residues of $\tilde{\chi}(u)$ vanish at $u=u_{j} + 
\eta$ leads to the following Bethe equations 
\begin{align}
\MoveEqLeft \prod_{l=1}^{N} \frac{\sinh(u_{j} -\theta_{l} + 
\eta)}{\sinh(u_{j} -\theta_{l} - \eta)}
\frac{\sinh(u_{j} + \theta_{l} + 
\eta)}{\sinh(u_{j} + \theta_{l} - \eta)} \non \\
& = 
\prod_{k=1;\, k \ne j}^{m} 
\frac{\cosh(\tfrac{1}{2}(u_{j} - u_{k}) + \eta)}
{\cosh(\tfrac{1}{2}(u_{j} - u_{k}) - \eta)}
\frac{\cosh(\tfrac{1}{2}(u_{j} + u_{k}) + \eta)}
{\cosh(\tfrac{1}{2}(u_{j} + u_{k}) - \eta)} \,, \non \\
& \qquad\qquad\qquad\qquad\qquad\qquad\qquad\qquad j = 1, \ldots, m\,.
\label{BEII}
\end{align}
These Bethe equations are unusual, as they involve $\cosh$ instead of 
$\sinh$ on the RHS. (For the $\varepsilon=0$ case 
\cite{Martins:2000xie, Nepomechie:2017hgw, Nepomechie:2018nvl}, the
Bethe equations are the same as (\ref{BEII}) except with
$\sinh$ on the RHS.)

The ansatz (\ref{tfactored})-(\ref{BEII}) is correct for $m=1$.  Indeed, we
prove it in Appendix \ref{sec:ABA}, and we have confirmed numerically that this ansatz
with $m=1$ correctly describes eigenvalues with even degeneracy for $N =2$
(degeneracy 6), $N =3$ (degeneracy 10) and $N =4$ (degeneracy 14).  We
expect that, for general $N$, this ansatz with $m=1$ describes eigenvalues
with degeneracy $4N-2$.

We also find numerically that this ansatz with $m=3$ describes the
eigenvalue for $N =3$ with degeneracy 2.  Unfortunately, we have not
succeeded to find more examples of even-degeneracy eigenvalues with $m>1$.
Hence, it appears that this ansatz cannot account for all the missing
eigenvalues.

\section{Conclusions}\label{sec:conclusion}

One of the aims of this paper is to draw attention to the unexpected
difficulty in solving the integrable quantum-group invariant
$D_{n+1}^{(2)}$ spin chains with $\varepsilon=1$.  We focused for
simplicity on the case $n=1$.  Using standard
assumptions, we arrived at a Bethe ansatz solution (\ref{factored})-(\ref{BE})
that is not complete. Indeed, the argument in Sec. \ref{subsec:conjecture} could
be regarded as a ``no-go theorem'', which we hope will motivate 
others to find a better approach.

We believe that we did succeed to describe a subset of the
transfer-matrix eigenvalues, namely, those with odd degeneracy.  The
remarkably simple solution (\ref{square})-(\ref{BEsimple}) 
suggests that there may be a connection to the XXZ model. 
Unfortunately, as $N$ increases, the fraction of eigenvalues with odd 
degeneracy rapidly decreases. 

The ``missing'' eigenvalues (namely, those with even degeneracy) may 
be related to a higher symmetry of the transfer matrix, as discussed 
in Sec. \ref{subsubsec:degensym}. It would be interesting to also
understand this symmetry better. The ansatz 
(\ref{tfactored})-(\ref{BEII}) may also provide a hint about 
the eventual complete solution.

\section*{Acknowledgments}
We thank Niall Robertson and Hubert Saleur for encouragement, and 
Nicolas Cramp\'e for discussions.  RN
also thanks Wen-Li Yang for valuable discussions, and for his warm
hospitality at the Institute of Modern Physics at Northwest University
in Xian. AR also thanks Marius de Leeuw, Anton Pribytok and Paul Ryan 
for helpful discussions.
RN was supported in part by the Chinese Academy of Sciences
President's International Fellowship Initiative Grant No.
2018VMA0017, and by a Cooper fellowship.  AR was supported by the
S\~ao Paulo Research Foundation FAPESP under the process \#
2017/03072-3 and \# 2015/00025-9. RP thanks the Institut Denis Poisson for hospitality
and the support from FAPESP and the Coordination for the Improvement of Higher Education Personnel (CAPES), process \# 2017/02987-8 and \#88881.171877/2018-01

% \newpage
% \clearpage

\begin{table}[h!]
\centering
\begin{tabular}{|c|c|c|}
\hline
degeneracy & multiplicity & $m$ \\   
\hline
3 & 1  & 0 \\
\hline
1 & 1  & 2 \\
\hline
\end{tabular}
\caption{\small $N=1$}\label{table:N1}
\end{table}

\begin{table}[h!]
\centering
\begin{tabular}{|c|c|c|}
\hline
degeneracy & multiplicity & $m$ \\   
\hline
5 & 1  & 0 \\
\hline
1 & 2  & 2 \\
\hline
3 & 1  & 4 \\
\hline
6 & 1  & ? \\
\hline
\end{tabular}
\caption{\small $N=2$}\label{table:N2}
\end{table}

\begin{table}[h!]
\centering
\begin{tabular}{|c|c|c|}
\hline
degeneracy & multiplicity & $m$ \\   
\hline
7 & 1  & 0 \\
\hline
3 & 3  & 2 \\
\hline
1 & 3  & 4 \\
\hline
5 & 1  & 6 \\
\hline
10 & 2  & ? \\
\hline
6 & 3  & ? \\
\hline
2 & 1  & ? \\
\hline
\end{tabular}
\caption{\small $N=3$}\label{table:N3}
\end{table}

\begin{table}[h!]
\centering
\begin{tabular}{|c|c|c|}
\hline
degeneracy & multiplicity & $m$ \\   
\hline
9 & 1  & 0 \\
\hline
5 & 4  & 2 \\
\hline
1 & 6  & 4 \\
\hline
3 & 4  & 6 \\
\hline
7 & 1  & 8 \\
\hline
14 & 3  & ? \\
\hline
10 & 8  & ? \\
\hline
6 & 12  & ? \\
\hline
2 & 4   & ? \\
\hline
\end{tabular}
\caption{\small $N=4$}\label{table:N4}
\end{table}

\begin{table}[h!]
\centering
\begin{tabular}{|c|c|c|}
\hline
degeneracy & multiplicity & $m$ \\   
\hline
11 & 1  & 0 \\
\hline
7 & 5  & 2 \\
\hline
3 & 10  & 4 \\
\hline
1 & 10  & 6 \\
\hline
5 & 5  & 8 \\
\hline
9 & 1  & 10 \\
\hline
18 & 4  & ? \\
\hline
14 & 15  & ? \\
\hline
10 & 35  & ? \\
\hline
6 & 40   & ? \\
\hline
2 & 16   & ? \\
\hline
\end{tabular}
\caption{\small $N=5$}\label{table:N5}
\end{table}

\newpage

\appendix

\section{First steps of the algebraic Bethe ansatz}\label{sec:ABA}

In this appendix we construct 1-particle states of the transfer matrix (\ref{transfer})
by means of the algebraic Bethe ansatz. We restrict our attention to the case $\varepsilon=1$ with 
$p=0$; and, for simplicity, we set the inhomogeneities to 
zero, $\theta_{j}=0$. The results are consistent 
with the $m=1$ case of the ansatz (\ref{tAnsatz}). 

We start by setting the following representation for the double-row 
monodromy matrix in the auxiliary space
\be
U_{a}(u) = T_a(u)\,  K^{R}_{a}(u)\, 
\widehat{T}_a(u)  =\left(
\begin{array}{cccc}
\mD_1(u) & \mB_1(u) & \mB_2(u) & \mB_3(u) \\
\mC_1(u) & \mA_2(u) & \mB_4(u) & \mB_5(u)\\
\mC_2(u) & \mC_4(u) & \mA_3(u) & \mB_6(u)\\
\mC_3(u) & \mC_5(u) & \mC_6(u) & \mA_4(u)\\
\end{array}
\right)_a\,,
\label{repU}
\ee
whose elements satisfy the reflection algebra
\be
R_{12}(u - v)\, U_1(u)\ R_{21} (u + v)\, U_2(v)
= U_2(v)\, R_{12}(u + v)\, U_1(u)\, R_{21}(u - v)  \,.
\label{reflecalg}
\ee
It is convenient to also define the following double-row operators
\begin{align}
\mD_j(u) &= \mathcal{A}_j(u) + \frac{ e^u \cosh (u) \sinh (2 \eta )}{\sinh (2 (u-\eta ))}\mD_1(u)\,,\quad \textrm{for} \quad j=2,3\,,\non\\
\mD_4(u) &= \mathcal{A}_4(u) - \frac{4 e^{2 u} \cosh (2 u-\eta ) \cosh (\eta ) \sinh ^2(\eta )}{\sinh (2 (u-\eta ))^2}\mD_1(u)
+\frac{e^u \cosh (\eta )}{\cosh (u-\eta )}(\mD_2(u)+\mD_3(u))\,,\non\\
\mB(u) &= \mB_4(u) + \frac{e^u \sinh (u) \sinh (2 \eta )}{\sinh (2 (u-\eta ))}\mD_1 (u)
- \frac{1}{2}\coth (\eta ) (\mD_2(u)+\mD_3(u))\,,\non\\
\mC(u) &= \mC_4(u) + \frac{e^u \sinh (u) \sinh (2 \eta )}{\sinh (2 (u-\eta ))}\mD_1 (u)
- \frac{1}{2} \coth (\eta )(\mD_2(u)+\mD_3(u))\,.
\end{align}
From the form of the R and K-matrices, it follows that the action of the double-row operators on the reference state
$\label{refstate}
\refs=\left( 
\begin{array}{c}
1 \\ 
0 \\
0 \\
0
\end{array}
\right)^{\otimes N}
$
is given by
\be\label{vac}
\mD_i(u)\refs = \Lambda_i(u) \refs \,,\qquad
\mB(u)\refs=\mC(u)\refs = 0\,,
\qquad \mC_j(u)\refs=0\quad \textrm{for} \quad j\neq 4 \,,
\ee
where
\begin{align}
\Lambda_1(u) &=-\frac{\sinh (u-\eta )}{\sinh (u+\eta )}16^N\sinh(u-2\eta)^{4N}\,,\\
\Lambda_2(u) &=\Lambda_3(u)=-\frac{e^{\eta } \sinh (2 u) \sinh (\eta )}{2 \cosh (u-\eta ) \sinh (u+\eta )}16^N\sinh^{2N}(u)\sinh^{2N}(u-2\eta)\,,\\
\Lambda_4(u) &=-\frac{e^{2 \eta } \cosh ^2(u) \sinh (u) \sinh (u-2 \eta )}{\cosh (u-\eta )^2 \sinh (u-\eta
   ) \sinh (u+\eta )}16^N\sinh^{4N}(u)\,.
\end{align}
It follows from (\ref{repU}) that the transfer matrix 
$t(u) = \tr_{a} K_{a}^{L}(u) U_{a}(u)$ can be written as 
\begin{align}\label{transferU}
t(u) &= -\frac{\cosh ^2(u-2 \eta ) \sinh (u) \sinh (u-2 \eta )}{\cosh (u-\eta )^2 \sinh (u-3 \eta )
   \sinh (u-\eta )}\mD_1(u) - \frac{e^{-2 \eta } \sinh (u-\eta )}{\sinh (u-3 \eta )}\mD_4(u)\non\\
&-\frac{e^{-\eta } \sinh (2 (u-2 \eta ))}{2 \cosh (u-\eta ) \sinh (u-3 \eta ) \sinh (\eta )}(\mD_2(u)+\mD_3(u))
\non\\
&-\frac{\cosh (\eta ) \sinh (u-2 \eta )}{\sinh (u-3 \eta )}(\mB(u)+\mC(u))\,,
\end{align}
from which it follows, using (\ref{vac}), that $\refs$ is an
eigenstate of the transfer matrix with eigenvalue given by $m=0$ in
(\ref{Ansatz}) or (\ref{tAnsatz}).  

Having settled the reference
state, the next step is to identify a 1-particle creation 
operator.  We find that $\mB_1(v)$ is suitable for that purpose.  
Indeed, we can use the reflection algebra (\ref{reflecalg}) to obtain the following
relations\footnote{Specifically, we use the following matrix 
elements of the reflection algebra (\ref{reflecalg}): from the entry $(1,5)$, we obtain $\mD_1(u)\mB_1(v)\refs$. Next, we use entries $(1,9)$ and $(5,13)$ to extract $\mD_1(u)\mB_2(v)\refs$ and $\mD_1(u)\mB_5(v)\refs$, respectively. This last step allows us to obtain $\mD_2(u)\mB_1(v)\refs$ from $(5,6)$. From the entry $(9,13)$, we obtain $\mD_1(u)\mB_6(v)\refs$, and then $\mD_3(u)\mB_1(v)\refs$ from $(9,10)$. Finally, we obtain $\mB(u)\mB_1(v)\refs$, $\mC(u)\mB_1(v)\refs$ and
$\mD_4(u)\mB_1(v)\refs$ from the entries $(5,10)$, $(9,6)$ and $(13,14)$, respectively.}
\begin{align}
\mD_i(u)\mB_1(v) \refs &= f_i(u,v)\Lambda_i(u)\mB_1(v)\refs + 
\textrm{unwanted}\,, \label{commutD} \\
\mB(u)\mB_1(v) \refs & = g(u,v)\Lambda_2(u)\mB_1(v)\refs + 
\textrm{unwanted}\,,  \label{commutB}\\
\mC(u)\mB_1(v) \refs &= -g(u,v)\Lambda_2(u)\mB_1(v)\refs + 
\textrm{unwanted}\,, \label{commutC}
\end{align}
where 
\begin{align}
f_1(u,v) &=\frac{\sinh (u+v) \sinh (u-v+2 \eta )}{\sinh (u-v) \sinh (u+v-2 \eta )}\,,\non\\
f_2(u,v) &=f_3(u,v)=\frac{\cosh (2 (u-\eta ))-\cosh (2 (v-\eta ))+1-\cosh (4 \eta )}{2 \sinh (u-v) \sinh (u+v-2\eta)}\,,
\non\\
f_4(u,v) &=\frac{\sinh (u+v-4 \eta ) \sinh (u-v-2 \eta )}{\sinh (u-v) \sinh (u+v-2 \eta )}\,,
\non\\
g(u,v) &=\frac{4 \cosh (u-\eta ) \cosh ^2(\eta ) \sinh (v-\eta )}{\sinh (u-v) \sinh (u+v-2 \eta )}\,.
\end{align}
Acting with the transfer matrix (\ref{transferU}) on $\mB_1(v)\refs$ and using 
the relations (\ref{commutD})-(\ref{commutC}), we obtain the $m=1$ off-shell 
equation
\begin{align}
t(u)\mB_1(v)\refs &=\Lambda(u)\mB_1(v) \refs \\ 
&\hspace{-1.0cm}+E(v)\left(F_1(u,v) \mB_1(u) \refs + F_2(u,v)\mB_2(u) \refs 
+ F_5(u,v)\mB_5(u) \refs + F_6(u,v)\mB_6(u) \refs \non\right) \,,
\end{align}
where
\begin{align}
\label{Lam1p}
\Lambda(u)&=-\frac{\cosh ^2(u-2 \eta ) \sinh (u) \sinh (u-2 \eta )}{\cosh (u-\eta )^2 \sinh (u-3 \eta )
   \sinh (u-\eta )}\Lambda_1(u)f_1(u,v)\\
&-\frac{e^{-2 \eta } \sinh (u-\eta )}{\sinh (u-3 \eta )}\Lambda_4(u)f_4(u,v)-\frac{e^{-\eta } \sinh (2(u-2 \eta) )}{\cosh (u-\eta ) \sinh (u-3 \eta )
   \sinh (\eta )}\Lambda_2(u)f_2(u,v)\,,\non
\end{align}
and
\begin{align}
E(v) & =\Lambda _2(v) \sinh (2 (v-\eta ))-e^{\eta } \sinh (\eta ) 
\Lambda _1(v) \sinh (2 v)\,, \\
F_1(u,v) &=-\frac{e^{-2 \eta } \cosh (\eta ) \sinh (2 (u-2 \eta ))}{2 \sinh (u-v) \sinh (u-3 \eta )
   \sinh (u+v-2 \eta ) \sinh (2 (v-\eta ))}\non\\
&\times \Big[
\cosh (u)+\cosh (v)-2 \cosh (u-2 \eta )-\cosh (v-2 \eta )+\cosh (u+2 \eta )
\non\\
&\quad
+4\cosh(\tfrac{1}{2}(u-v+2\eta)) \sinh(\tfrac{1}{2}(u+v)) \cosh(\eta)
\Big] \,, \\
F_2(u,v) &=
-\frac{e^{-2 \eta } \cosh (\eta ) \sinh (2 (u-2 \eta ))}{2 \sinh (u-v) \sinh (u-3 \eta )
   \sinh (u+v-2 \eta ) \sinh (2 (v-\eta ))}
\non\\
&\times \Big[
\cosh (u)-\cosh (v)-2 \cosh (u-2 \eta )+\cosh (v-2 \eta )+\cosh (u+2 \eta )
\non\\
&\quad
+4\sinh(\tfrac{1}{2}(u-v+2\eta)) \cosh(\tfrac{1}{2}(u+v)) \cosh(\eta)
\Big] \,, \\
F_5(u,v) &=\frac{e^{-2 \eta } \cosh (\eta ) \sinh (2 (u-2 \eta ))}{2 \cosh \left(\frac{1}{2} (u+v-2
   \eta )\right) \sinh \left(\frac{1}{2}(u-v)\right) \sinh (u-3 \eta ) 
   \sinh (2 (v-\eta ))}\,, \\
F_6(u,v) &=\frac{e^{-2 \eta } \cosh (\eta ) \sinh (2 (u-2 \eta 
))}{2\sinh \left(\frac{1}{2} (u+v-2 \eta )\right) \cosh 
\left(\frac{1}{2}(u-v)\right)
   \sinh (u-3 \eta )  \sinh (2 (v-\eta ))}\,.
\end{align}

Setting $v=u_1+\eta$, we identify the first and second 
terms in (\ref{Lam1p}) (involving $\Lambda_1(u)$ and $\Lambda_4(u)$) 
as $\phi(u)Z_1(u)$ and $\phi(u)Z_4(u)$ in (\ref{tAnsatz}), respectively. Moreover, by noticing that for $m=1$ the function $f_2(u,v)$ can be written as
\be
f_2(u,v) = 
\frac{Q(u-\eta)Q(u-\eta+i\pi)+\sinh^2(\eta)\cosh^2(\eta)}{Q(u-\eta)Q(u-\eta+i\pi)}\,,
\ee
as well as the identity (valid for $m=1$)
\be
&&Q(u-3\eta)Q(u+\eta+i\pi)+Q(u+\eta)Q(u-3\eta+i\pi)
\non\\&&\quad=
2Q(u-\eta)Q(u-\eta+i\pi)-2\sinh^2(\eta)\cosh^2(\eta)\,,
\ee
we conclude that the T-Q equation 
(\ref{Lam1p}) is consistent with the ansatz (\ref{tAnsatz}) for $m=1$. In addition,
we see that the Bethe equations (\ref{BEII}) for $m=1$ correspond to 
one of the possible branches of $E(u_1+\eta)=0$. We also 
remark that (\ref{Lam1p}) has the form of an inhomogeneous T-Q 
equation.

\newpage
\clearpage

% \bibliographystyle{utphys}
% \bibliography{refs}

\providecommand{\href}[2]{#2}\begingroup\raggedright\endgroup

\end{document}